\documentclass{article}

\usepackage{graphicx}

\def\abstract#1{\vskip 7mm 
        \begin{center}{\large Abstract}\par \smallskip
                \begin{minipage}[c]{12cm}
                        \small #1
                \end{minipage}
        \end{center}
}
\def\title#1{\begin{center}{\Large\bf #1}\end{center}}
\def\author#1{\vskip 5mm \begin{center}{#1}\end{center}}
\def\address#1{\begin{center}{\it #1}\end{center}}
\makeatletter
\@ifundefined{lesssim}{}{}
\@ifundefined{gtrsim}{}{}
\def\vereq#1#2{\lower3pt\vbox{\baselineskip1.5pt \lineskip1.5pt
\ialign{$\m@th#1\hfill##\hfil$\crcr#2\crcr\sim\crcr}}}
\makeatother

\begin{document}

\title{%
  General Relativity and Gravitation
  \smallskip \\
  {\large --- Millisecond oscillators in accreting neutron stars and
     black holes  ---}
}
\author{%
  W{\l}odek Klu\'zniak,\footnote{On leave of absence from the Copernicus 
   Astronomical Center. E-mail:wlodek@camk.edu.pl}
  Marek Abramowicz\footnote{E-mail:marek@fy.chalmers.se}
}
\address{%
  Institute of Astronomy, Zielona G\'ora University, ul. Lubuska 2,
   65 265 Zielona G\'ora, Poland\\
   CESR, 9 ave. Colonel-Roche, 31028 Toulouse, Cedex 4, France\\
  Astrophysics Department, Chalmers University, S-41296 G\"oteborg, Sweden
}

\abstract{
  Millisecond variability detected in low-mass X-ray binaries (LMXBs)
probably reflects motions of accreting fluid in the strong-field
gravity of neutron stars or neutron stars. Parametric resonance
between two oscillators with unequal eigenfrequencies
provides a natural explanation for the 2:3 frequency ratio
observed in some black-hole systems, and probably also in the brightest
neutron stars. The two oscillators likely correspond
to the meridional and radial epicyclic motions.
}

\section{Millisecond variability in LMXBs}

Frequencies as large as 1.2 kHz in the X-ray flux of LMXBs have been detected in 1996 by the first large satellite detector to have the requisite time resolution,
the Rossi X-ray Timing Explorer.
For a review of observations see van der Klis 2000, and for that of the
previously discovered lower frequency variability, van der Klis 1989.

Millisecond variability of X-ray emission from accreting neutron stars
and black holes has been expected at least from the time of
Shvartsman's 1971 publication. Bath (1973) pointed out that in
cataclysmic variables
a natural timescale of variation is given by the Keplerian
frequency and that clumps orbiting in the accretion disk could modulate the flux.  For stellar-mass black holes and for neutron stars 
such clumps would induce millisecond variability of the X-ray flux
(Sunyaev 1973).
After quasi-periodic signals (QPOs) had been observed at frequencies corresponding 
to Keplerian motion in the inner parts of the accretion disk in
a strongly magnetized accreting neutron star, the X-ray pulsar
EXO 2030+375 (Angelini et al. 1989),
Klu\'zniak, Michelson and Wagoner (1990) pointed out that scaling
to the case of a moderately rotating compact neutron star of negligible
magnetic field would yield a QPO at the frequency of orbital motion in the marginally stable orbit,
$f_K=2.2 \,{\rm kHz}(1+0.75j)M_\odot/M$
(with $j$ the dimensionless angular momentum of the star), and that
this frequency, when observed, 
could be used to obtain a direct estimate of the stellar mass $M$.

Another line of reasoning originates with Okazaki, Kato and Fukue (1987),
who pointed out that the radial epicyclic frequency 
in the space time metric around a black hole (or a neutron star)
has a maximum, and goes to zero in the marginally stable orbit,
which allows certain modes of vibration to be trapped
in the inner region of the accretion disk.
Diskoseismology, the theory of vibrations of accretion disks, has been developed
in detail in the following years and has been recently reviewed
by Wagoner (1999) and Kato (2001). In particular, it predicts
the appearance of certain stable frequencies of quasi-periodic motion
in black-hole accretion disks. The fundamental g-mode and c-mode,
especially, have frequencies which are a function of the black-hole 
mass and the Kerr parameter $a$, alone, and which agree well with those
observed in the micro-quasars.

The kHz QPOs observed in accreting neutron stars are more difficult to
understand, as the observed frequencies vary considerably on a timescale
of minutes or hours. Further, typically, two high frequency peaks
in the power spectra are observed, where only one had at first been reported in black holes. It has been thought, initially, that
the separation between the two peaks corresponds to the spin frequency
of the neutron star but, in fact, this difference frequency is not
constant either (e.g., van der Klis et al. 1997), whereas the spin
frequency can vary only on the accretion time-scale, which is 
millions of years in LMXBs (even in the brightest sources the accretion rate is less that $10^{-7}M_\odot$/y, and the neutron star mass is on
the order of $1M_\odot$).

In spite of these differences, the power spectra of both the neutron star sources and the black hole ones are quite similar, the disparity in the characteristic frequencies being easy to explain by the higher mass of the
black holes. Also, in the neutron star sources, the maximum frequency
observed varies remarkably little with the accretion rate, or the
type of the source. All this points to an accretion disk origin of the phenomenon. 

\begin{figure}
\centering
\includegraphics[width=7.0cm]{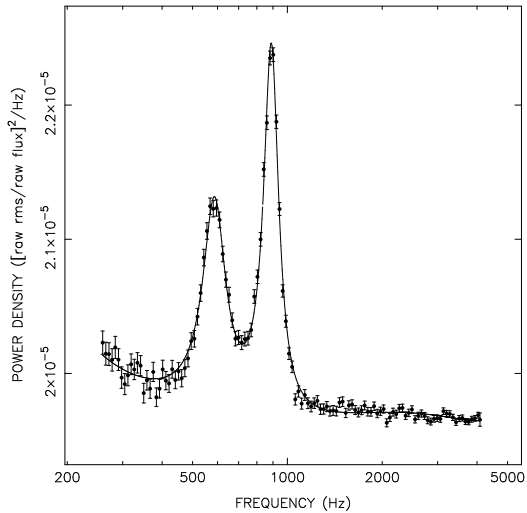}
\caption{The two ``kHz" QPOs of Sco X-1. The figure is from van der Klis et al. 1997.}
\label{}
\end{figure}
\begin{figure}
\centering
\includegraphics[width=10.0cm]{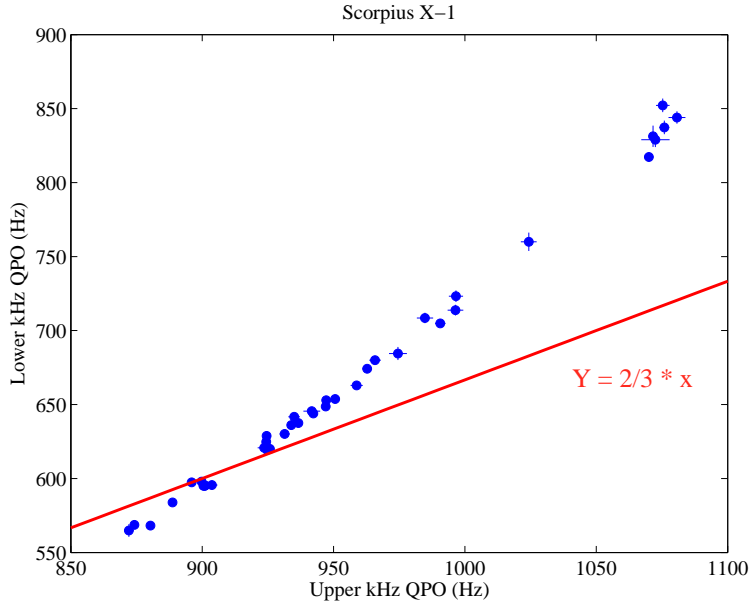}
\caption{The frequencies of the two kHz QPOs in Sco X-1 vary with time,
but remain correlated. Neither the difference, nor the ratio,
of the two frequencies remains constant. The data is from van der Klis et al. 1997. Also shown is a line of constant 
frequency ratio (we thank an anonymous referee of Astronomy and Astrophysics for kindly plotting the data).
}
\label{}
\end{figure}
\section{The accretion disk in LMXBs}

There is little doubt that mass transfer from the binary companion
onto a neutron star or a black hole is responsible for the observed
X-ray emission in low-mass X-ray binaries 
(Shklovsky 1967, Zel'dovich and Guseynov 1966).
Although the central region around the compact object is too small
to be resolved with current instruments (being only several tens
of kilometers across, observed from a distance of kiloparsecs)
it is generally agreed that the accreting fluid forms a disk
around the central compact object (Prendergast and Burbidge 1968). Whether this is
a geometrically thin, and hence rotation-supported, accretion
disk (Shakura and Sunyaev 1973), or a sub-Keplerian,
geometrically thick disk (Jaroszy\'nski et al. 1980), is a matter of some dispute.
For an artist's impression of the general appearance of the system
see, e.g., the cover of the book by Kato, Fukue and Mineshige (1998).

There are two analogous types of systems in which the disk can
be resolved, and in these the disk is geometrically thin.
These are cataclysmic variables (binaries in which the accreting
object is a non-magnetic white dwarf), and the celebrated extragalactic
quasars. The correspondence between the physical processes in
quasars and the Galactic X-ray novae, dubbed micro-quasars,
has already been remarked upon in the literature 
(e.g., Mirabel and Rodriguez 1998),
the main difference between the real and the ``micro" quasars
being in the mass of the black hole: millions or billions of solar
masses as opposed to about ten $M_\odot$.

Water-vapour masers have been observed in some of the quasars,
and radio observations allowed the mapping of the geometrical
distribution of the emitters, as well as of their velocity distribution.
The results leave no doubt---in NGC 4258 the masers are present
in a geometrically thin and warped disk, and they orbit the central
source with Keplerian velocities (Miyoshi et al. 1995).
Perhaps the accretion disks in LMXBs are also Keplerian,
although most of what we have to say here will apply equally well
to some accretion tori.

\section{Two variable kHz frequencies: non-linear resonance?}

If a common framework can be found
for understanding the rapid variability of the X-ray flux in both
the black hole and the neutron star sources in LMXBs,
the behavior of QPOs in the latter may offer the key to unlocking the secrets of the former.

The most characteristic property of kHz QPOs in neutron stars
is that there are typically two peaks in the power spectrum,
located at frequencies which vary in time, but only within a limited, characteristic range (Fig.~2).
A single variable
frequency can be identified already in the Newtonian regime.
It is also easy to find two characteristic constant frequencies. But we would like to ask what may be the origin of 
{\sl two, characteristic}, but
{\sl variable} frequencies, in nearly Keplerian motion around a neutron star.

\begin{figure}
\centering
\includegraphics[width=8.5cm]{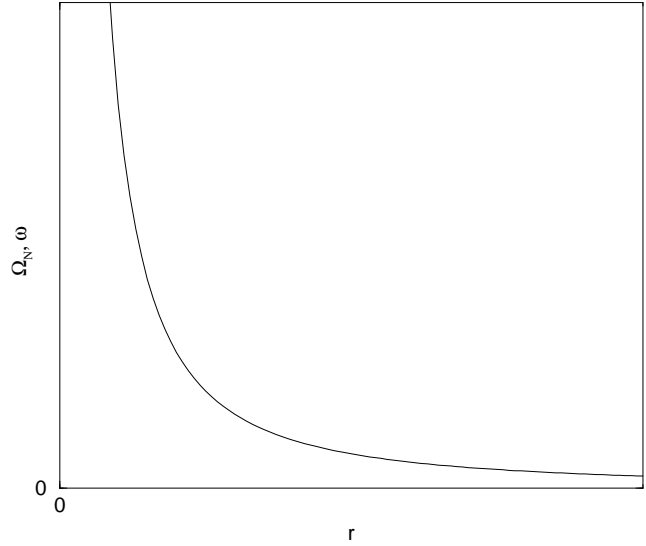} 
\caption{The orbital and epicyclic frequencies for a Newtonian
$1/r$ potential. Note that there are no preferred frequencies,
the problem is scale-free.}
\label{sco}
\end{figure}
\begin{figure}
\centering
\includegraphics[width=8.5cm]{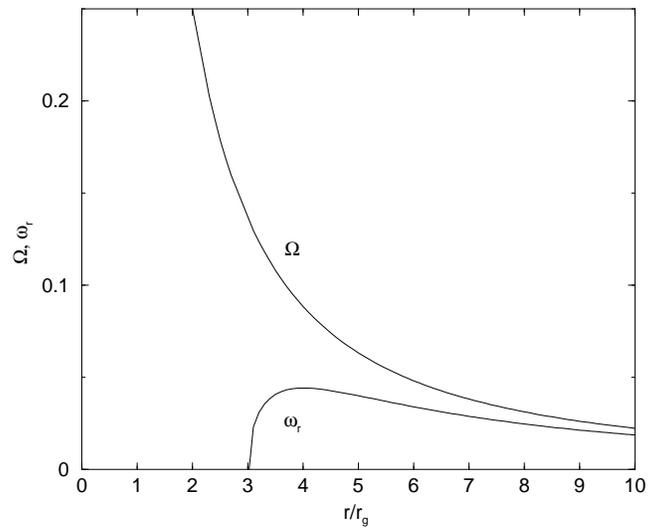}
\caption{The orbital and the radial epicyclic frequencies in 
the Schwarzschild metric (after Kato et al. 1998). The Schwarzschild
radius $r_g\equiv 2GM/c^2$ sets a length-scale and the corresponding
characteristic frequencies.}
\label{}
\end{figure}

In Newtonian gravity of a spherically symmetric point mass,
there is no preferred frequency (Fig.~3). Further, there is degeneracy
between the orbital frequency in circular and not-quite-circular
motion---this is why bound orbits are closed curves (Kepler's first law).
So if neutron star exteriors were described by this theory, one would expect
temporal fluctuations on all time-scales in the accretion disk,
corresponding to the (huge) range of radii covered by the accreting
fluid, but no characteristic frequencies.

However, 
characteristic frequencies can be induced in a Keplerian distribution
externally---the gaps in the rings of Saturn are a well-known example.
The radius of a neutron star introduces a characteristic length-scale
to the problem, and a corresponding frequency, the Keplerian
frequency at the surface of the star. This frequency is fixed for
a given star, of course, and together with the spin frequency
(which, in LMXBS, is practically constant on a human time-scale)
allows two {\it fixed} frequencies to be identified in the Newtonian
problem of accretion onto a neutron star.

If the accretion disk is terminated by
the magnetosphere of the neutron star, its inner radius and the
corresponding orbital frequency can vary somewhat with the accretion rate.
This would lead to the appearance of a {\sl single} characteristic,
and variable, frequency, in addition to the constant stellar spin 
frequency. As an example take the X-ray pulsar
EXO 2030+375, already mentioned in Section 1:
the power spectrum of its X-ray flux exhibits prominent harmonics
of the rotational frequency of the neutron star, in addition to
the $\sim 20$ mHz QPO with frequency varying in step with the luminosity
 (Angelini et al. 1989).

In general relativity orbital degeneracy is broken, bound orbits are no longer closed curves (``precession of the perihelion"), the frequency
of epicyclic motion induced by a radial perturbation differs from orbital
frequency. There is a preferred length-scale, the gravitational radius
$M(G/c^2)$, and corresponding characteristic frequencies: the radial epicyclic frequency goes to zero
at the innermost (marginally) bound circular orbit, and has a maximum
at a slightly larger orbit (Fig.~4). However, these are fixed frequencies.

In any sufficiently complicated system, resonances appear at characteristic frequencies. Of particular interest are resonances in non-linear systems,
these occur not only when the driving frequency is close
to the eigenfrequency of the oscillator, but also when the two frequencies
are (nearly) in a rational ratio, $m:n$, with the strongest resonances
occurring at the lowest values of the integers (Landau and Lifshitz 1976).
The appearance of two, somewhat variable, frequencies in the rather
complex accreting flow in LMXBs is strongly suggestive of a non-linear
resonance. 

\section{Epicyclic resonance?}

We note that there are two types of fluid perturbations
in nearly circular three-dimensional flow which correspond to the
two epicyclic frequencies familiar from study of test-particle orbits---the
radial epicyclic frequency of planar motion, and the meridional
epicyclic frequency of slightly non-planar motion. For a spherically
symmetric gravitating body, the meridional epicyclic frequency is
equal to the orbital frequency of the unperturbed circular orbit.
For co-rotating orbits in  the Kerr metric the meridional epicyclic frequency is always lower than the orbital frequency and higher
than the radial epicyclic frequency (e.g., Wagoner 1999),
and a similar relation obtains for rapidly rotating neutron stars.
Thus, at characteristic radii, the ratio of the two epicyclic
frequencies is rational, allowing the possibility of a non-linear
resonance between the two types of fluid perturbations.
We have suggested that this is the origin of the two kHz QPOs
in neutron stars (Klu\'zniak and Abramowicz 2001a).

What would be the implications of this suggestion for the QPOs
observed in black hole systems? When we first argued for the
presence of a non-linear epicyclic resonance in LMXBs, only a single
high-frequency QPO had been observed in black hole systems.
The most obvious prediction following from our hypothesis
was that a second QPO should be present also in black holes,
and that the two frequencies should be in a rational ratio.

We are greatly encouraged that a second QPO has indeed been
discovered in GRO J1655-40 (which reportedly has a $6M_\odot$
black hole) and that the two simultaneously observed
frequencies have been reported to be 450 Hz and 300 Hz (Strohmayer 2001a), clearly in a 3:2 ratio (Abramowicz and Klu\'zniak 2001).
Another system, GRO J 1550-564 with a $10M_\odot$ black hole, in one of its states also shows two QPOs at the same frequency ratio, and the frequencies of 270 Hz and 180 Hz can be obtained by mass scaling from those of 1655-40,
implying that the Kerr parameter of the two holes may be nearly identical (Remillard et al. 2002).
Fig.~5 (reproduced from Remillard et al. 2002)
illustrates how these ideas may be used to constrain
the spin of the black hole, when the mass is known and the observed frequencies are identified with linear combinations of the orbital
frequency and the radial epicyclic frequency.

\begin{figure}
\centering
\includegraphics[width=12.0cm]{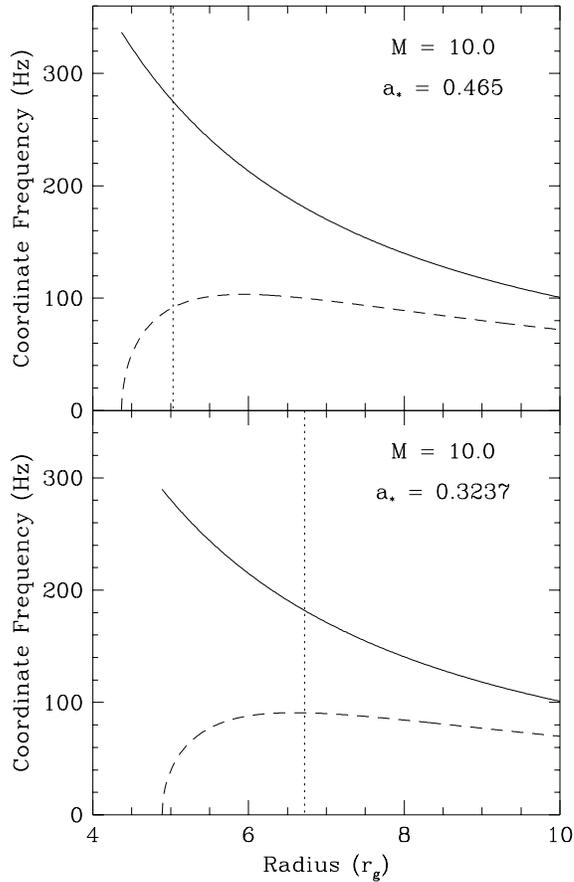}
\caption{This figure, reproduced from Remillard et al. 2002,
illustrates how the spin parameter of a black hole may be determined from
the resonant condition (Abramowicz and Klu\'zniak 2001).
The plot shows the orbital frequency $\Omega$ (continuous curve)
and the radial epicyclic frequency $\omega_r$ (dashed curve),
for two choices of the spin parameter for a $10M_\odot$ black hole.
The most prominent of the frequencies observed in XTE J1550-564 are 270 Hz and 180 Hz.
In the top panel they are identified with $\Omega$ and  
$\Omega-\omega_r$, in the bottom panel with $\Omega+\omega_r$ and 
$\Omega$, respectively. The vertical line indicates the radial position of
a 3:1 resonance between $\Omega$ and $\omega_r$ in the top panel,
and a 2:1 resonance in the lower panel 
(the radius is in units of $r_{\rm g}\equiv GM/c^2$).
In this contribution, we suggest that the actual resonance is a 2:3
parametric resonance between $\omega_r$ and the meridional epicyclic
frequency $\omega_z$ (not shown).}
\label{3}
\end{figure}

\begin{figure}
\centering
\includegraphics[width=10.0cm]{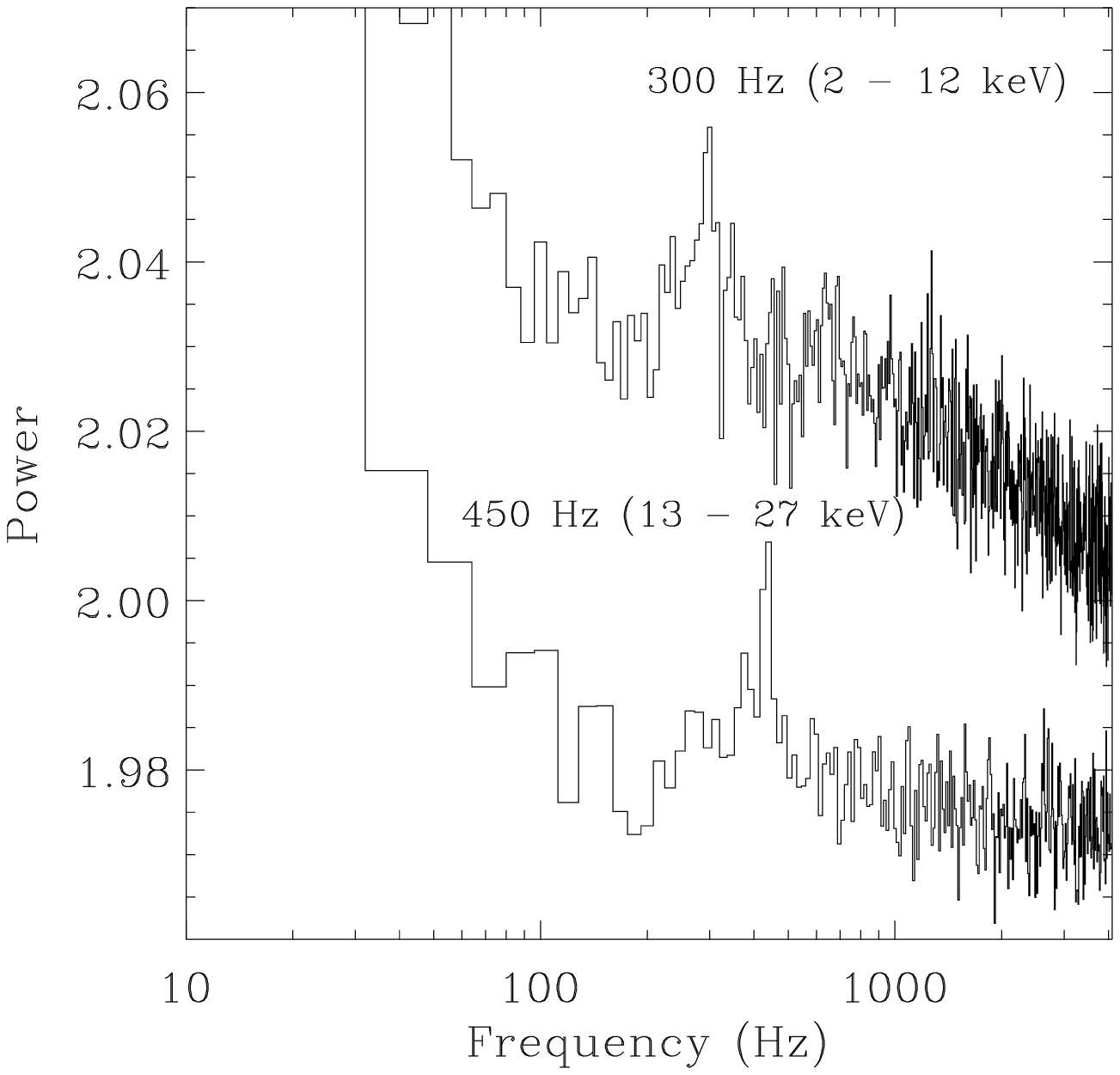}
\caption{The two high frequency QPOs detected in GRO J1655-40.
This figure is reproduced from Strohmayer 2001a.}
\label{4}
\end{figure}

\section{Parametric epicyclic resonance?}

We have seen that
the frequencies of these first twin QPOs to be discovered in black hole systems are in a 3:2 ratio. Why not 1:2 or 1:3\,?
There is a particular type of non-linear resonance in which
large-amplitude motions are easily excited by a small perturbation
of the right frequency. This is parametric resonance, in which
the eigenfrequency of an oscillator is varied periodically,
i.e., when the eigenfrequency of a system is of the form
$\omega=\omega_0(1+h\cos\omega_1 t)$, with $h<<1$.
The displacement of a harmonic oscillator with this eigenfrequency satisfies
the Mathieu equation.
Resonance occurs when 
$$\omega_1/\omega_0=2/n,$$
with $n$ integer;
the amplitude grows intially exponentially to a saturation value
determined by non-linearities,
the fastest growing modes correspond to the lowest values of the integer $n$ (Landau and Lifshitz 1976).
 For resonance to occur in the presence of dissipation, the size $h$ of the perturbation
has to exceed a certain threshold value.

One can write down the equations of disk perturbation in a form
in which the the vertical and radial components of motion
are coupled by a ``pressure" term proportional to the speed of sound squared. Typically, in a linear analysis, the term can
be neglected. But when a state of motion arises in which
this coupling is modulated at a frequency obeying the resonant
condition, this term becomes the most important one.
Specifically, there are states of disk motion in which
the dominant frequency is close to the radial epicyclic frequency
$\omega_r$, and other states of ``vertical" motion,
in which the frequency is close to the meridional epicyclic
frequency $\omega_z$. As we have seen, $\omega_r<\omega_z$
for black holes, always. Identifying $\omega_r$ with $\omega_1$
in the Mathieu equation, and $\omega_z$ with $\omega_0$, we see
that for parametric resonance to occur in the Kerr metric, both $$\omega_1<\omega_0$$
 and 
$$\omega_1/\omega_0=2/n$$
 must hold.
The lowest value of $n$ for which both eqs. are satisfied is
$n=3$. Hence, if 180 Hz is the value of the radial epicyclic frequency,
vertical motions at a frequency of 270 Hz are easily excited in the disk.

Incidentally, it is often assumed that accretion disks in these systems
have a hot corona. 
The observed energy dependence of the QPOs in GRO J1655-40
(Fig.~6) is easy to square with the notion that motions perpendicular to
the disk plane, i.e., the ones most likely to disturb the corona, occur predominantly at the higher frequency.

\section{Frequencies in the non-linear coupled system}

What is the expected harmonic content of the epicyclic oscillators
in resonance? First let us examine the non-linearities of the effective
potential. If the unperturbed disk is symmetric with respect to reflections in the equatorial plane, in agreement with the symmetry of the black hole, 
the anharmonicity of the meridional oscillator is of third order
in the equations of motion. In contrast, the equation of motion of the radial oscillator has second order anharmonic terms (the potential has cubic terms). This implies that subharmonic content is possible in the
motion of these oscillators (Landau and Lifshitz 1976).
The meridional oscillator may have a subharmonic at $\omega=(1/3)\omega_0$
and the radial oscillator a subharmonic at $\omega=(1/2)\omega_1$.
If the two oscillators are in a 2:3 resonance, $\omega_1=(2/3)\omega_0$, and the two subharmonics coincide!

It is noteworthy, that Remillard et al. (2002) report a third QPO
frequency in one of the states of XTE J1550-564. In addition to
270 Hz and 180 Hz, a weaker oscillation at 90 Hz is present (Fig.~5).
The third frequency coincides with the predicted subharmonic.

\section{1:2:3:5}

 If the two oscillators are in parametric resonance, as we have argued
above, the frequency of 90 Hz in XTE J1550-564
is predicted for a more direct reason. 
The Mathieu equation does not admit strictly harmonic solutions.
Instead, if the leading term in the solution is harmonic at frequency $\omega_0$, additional terms must be included at ``combination" frequencies, $\omega_0+\omega_1$ and $\omega_0-\omega_1$. For a 2:3 resonance these are at $(1/3)\omega_0$ and $(5/3)\omega_0$.

It is interesting to note that of these two combination frequencies
one has been detected in XTE J1550-564, while the other may  have been detected in GRS 1915+105.
Why this is the case may be difficult to understand.
Experience with asteroseismology shows that it is by far easier to
compute the spectrum of allowed frequencies than to predict
which of the frequencies will actually show up in the source at any given time.
Primarily this is because energy flow in the system is very sensitive
to the detailed, and poorly understood, coupling of modes in the non-linear system.
These difficulties are exacerbated in accretion disk theory,
which can offer only educated guesses as to the physical state
of the accreting fluid.

The state of affairs is this. The theory of parametric resonance
predicts a sequence of initial primes for the frequencies,
1:2:3:5, with 2 and 3 the most prominent.
The 2:3 have been detected in two sources, GRO 1655-40
(300 Hz : 450 Hz) and XTE J1550-564 (180 Hz : 270 Hz).
The ``1" (90 Hz) also in XTE J1550-564, 1:2:3 really (Fig.~7).
In a third system, 1915+105, the frequencies are 41.5 Hz and 69.2 Hz (Strohmayer 2001b), i.e., they are in a 3:5 ratio
(Klu\'zniak and Abramowicz 2002).

\begin{figure}
\centering
\includegraphics[width=8.5cm]{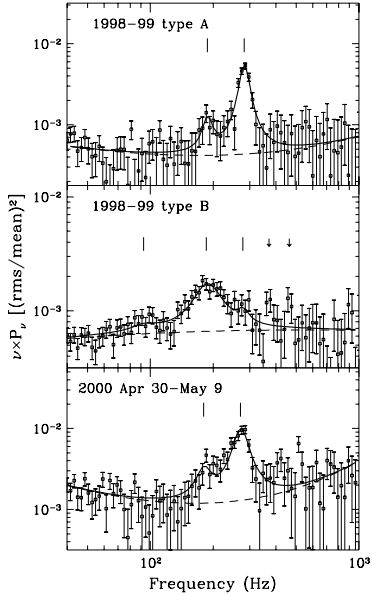}
\caption{This figure is reproduced from Remillard et al. 2002.
The panels show the power spectra in three different states of the
XTE J1550-564 system.
In all three, QPOs at 180 Hz and 270 Hz are clearly visible.
In the state named ``type B" (middle panel) a frequency of 90 Hz is also statistically
significant.}
\label{5}
\end{figure}

\section{Neutron star systems}

The observed ratios of QPO frequencies in black hole systems
agree very well with the theory of non-linear resonance,
and parametric resonance between epicyclic oscillators
in particular. We have been led to this theory by a search for the origin
of two variable QPO frequencies in neutron star systems,
and we view the subsequent discovery in black hole systems of double frequencies in rational ratios as confirmation of this theory.
However, in neutron stars, the ratio of the two frequencies is
not fixed (e.g., Fig.~2). Is the kHz QPO phenomenon in neutron stars and in black holes really a manifestation of the same physics?
If so, then a 2:3 ratio should also be discernible in LMXBs
containing neutron stars.

Let us begin by examining anecdotal evidence. In the van der Klis 2000
review, only two examples are shown of twin peaks in the power spectra.
We have reproduced the first one, showing the kHz QPOs of Sco X-1,  as Fig.~1. The second one, showing the twin QPOs of the neutron star source
4U 1608-52 (M\'endez et al. 1998), is reproduced here as Fig.~8.
Note that in both cases, the frequencies of the QPOs are close
to 600 Hz and 900 Hz, i.e., close to a 2:3 ratio. This could be coincidental, or it could reflect the fact that certain frequencies
are more likely to appear than others.
Abramowicz et al. (2002b) argue that the latter is the case, at least for Sco X-1.

For Sco X-1, the two frequencies seem to be scattered over a wide
range (Fig.~2), but the distribution of points along the line of
correlation is not random (Abramowicz et al. 2002b). One way to demonstrate this is to plot a histogram of the frequency ratios (Fig.~9).
It is striking that the observations cluster about the value 2/3.
We take this to be a very strong indication that a resonance
is involved in the simultaneous appearance of the two frequencies.
A corotation resonance has been considered (Kato 2002),
but it seems to be damped (Kato 2003a). Kato 2003b has shown that a parametric resonance may
excite both an axisymmetric and a non-axisymmetric disk (one-armed) modes whose frequencies are in a $1/\sqrt2\approx0.71$ ratio for the Schwarzschild metric.

Abramowicz et al. (2003) have investigated the parametric epicyclic
resonance for nearly circular and nearly geodesic motion.
Choosing an arbitrary, but fairly general form of coupling
between the radial and vertical components of motion, they have demonstrated the presence of parametric resonance, at frequencies
somewhat depending on the strength of coupling, which is taken to simulate
pressure effects in the disk. In resonance, the ratio of the frequencies
is accurately 2:3, but sizable amplitudes of motion are excited also
slightly off-resonance. Quantifying with a single adjustable parameter
how far off resonance the system is, the authors have reproduced
the slope of the frequency correlation found in Sco~X-1
(Fig.~10).

\begin{figure}
\centering
\includegraphics[width=7.0cm]{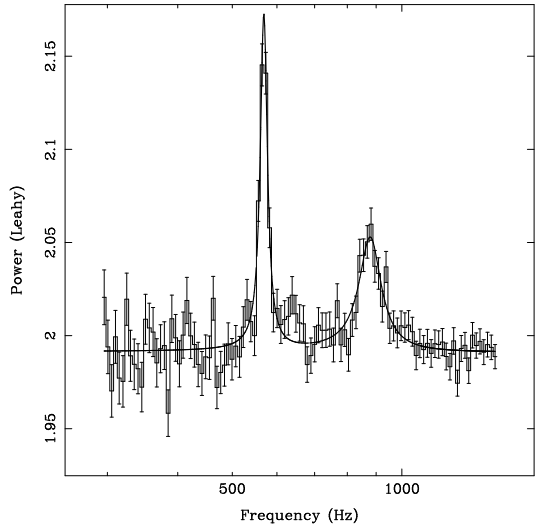}
\caption{The kHz QPOs of 4U 1608-52. The figure is from M\'endez et al. 1998. See also the review by van der Klis (2000).}
\label{}
\end{figure}
\begin{figure}
\centering
\includegraphics[width=8.5cm]{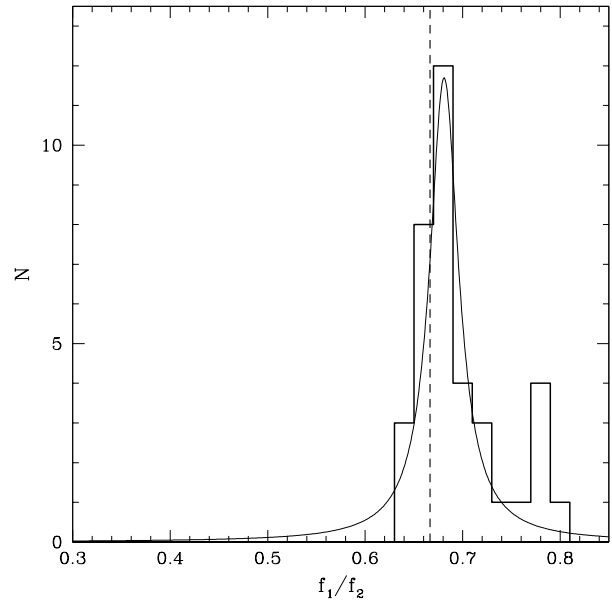}
\caption{A histogram of the ratio of frequencies in Sco X-1.
Also shown is the best-fit Lorentzian, the vertical dashed line
indicates where the frequencies are in a 2:3 ratio.
The figure is from Abramowicz et al. (2002b).}
\label{}
\end{figure}
\begin{figure}
\centering
\includegraphics[width=8.5cm]{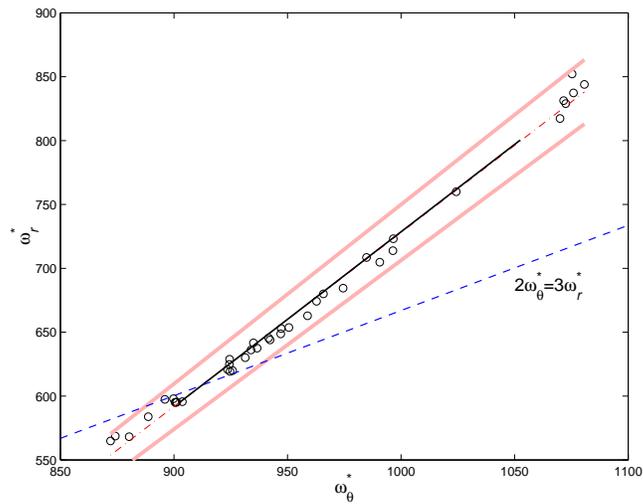}
\caption{Frequencies slightly off-resonance from a calculation
of parametric resonance in nearly-geodesic motion (thick line)
superposed on the Sco X-1 data (compare Fig.~2).
The figure is from Abramowicz et al. (2003).}
\label{}
\end{figure}


\begin{thebibliography}{99}

\bibitem{AK02}
Abramowicz M.A., Almergren G.J.E., Klu\'zniak W.,
 Thampan A.V., Wallinder F., 2002a, CQG 19, L57

\bibitem{ABBK02}
Abramowicz M.\,A., Bulik T., Bursa~M., Klu\'zniak~W., 2002b,
http://xxx.lanl.gov/abs/astro-ph/0206490

\bibitem{}
Abramowicz, M.A., Karas, V., Klu\'zniak, W., Lee, W.H., Rebusco, P.,
2003, PASJ, in press, astro-ph/0302183

\bibitem{}
Abramowicz, M.A., Klu\'zniak, W., 2001, Astron. Astrophys. 374, L19,

\bibitem{}
Angelini, L., Stella, L., Parmar A., 1989, ApJ 346, 906

\bibitem{}
Jaroszy\'nski, M., Abramowicz, M. A., Paczy\'nski, B., 1980,
Acta Astronomica, 30, 1.

\bibitem{}
Bath, G. T. 1973, Nature Ph.Sc. 246, 84

\bibitem{K01}
Kato S., 2001, PASJ 53, 1

\bibitem{}
Kato S., 2002, PASJ 54, 39

\bibitem{}
Kato S., 2003a, PASJ, in press

\bibitem{}
Kato S., 2003b, PASJ, in preparation

\bibitem{}
Kato, S., Mineshige, S., and Fukue, J. 1998, {\it Black Hole Accretion
Disks} (Kyoto University Press, Kyoto).

\bibitem{}
Klu\'zniak, W., Abramowicz, M.A., 2001a, astro-ph/0105057

\bibitem{KluzA01}
Klu{\'z}niak, W.,  M.A. Abramowicz 2001b,  Acta Phys. Pol. B, 32, 
3605, available at http://th-www.if.uj.edu.pl/acta/

\bibitem{}
Klu\'zniak, W., Abramowicz, M.A., 2002, astro-ph/0203314

\bibitem{KluznMW90}
Klu{\'z}niak, W., Michelson, P., Wagoner, R.V. 1990, ApJ 358, 538

\bibitem{LL76}
Landau, L.D., Lifshitz, E.M., 1976, {\it Mechanics} 
(Pergamon Press, Oxford) 

\bibitem{}
M\'endez, M., van der Klis, M., Wijnands, R., Ford, E., van Paradijs, J.,
Vaughan, B.A., 1998, ApJ 505, L23

\bibitem{}
Mirabel, I. F.,, Rodriguez, L. F. 1998, Nature 392, 673

\bibitem{}
Miyoshi, M., et al., 1995, Nature 373, 127

\bibitem{OKF87}
Okazaki A.\,T., Kato S., Fukue J., 1987, PASJ 39, 457

\bibitem{}
Prendergast, K. H.,  Burbidge, G. R.,  1968, ApJ 151, L83

\bibitem{}
Remillard R.~A., Muno M.P., McClintock J.E., Orosz J.A.,
2002, ApJ 580, 1030

\bibitem{}
Shakura, N. I., Sunyaev, R. A., 1973,  Astron. Astrophys.  24, 337

\bibitem{}
Shklovsky, I.S., 1967, ApJLett. 148, L1

\bibitem{}
Shvartsman, V.F., 1971, Soviet Astron. 15(3), 377

\bibitem{}
Strohmayer, T.E. 2001a, ApJ. 552, L49

\bibitem{}
Strohmayer, T.E. 2001b, ApJ. 554, L37

\bibitem{}
Sunyaev, R. 1973, Soviet Astron. 16(6), 941

\bibitem{}
van der Klis, M.  1989, Ann. Rev. Astron. Astrophys. 27, 517

\bibitem{}
van der Klis, M.  2000, Ann. Rev. Astron. Astrophys. 8, 717

\bibitem{}
van der Klis, M., Wijnands, R.A.D., Horne, K., Chen, W. 1997,
ApJ 481, L97

\bibitem{W99}
Wagoner R.W., 1999, Phys. Rev. 311, 259

\bibitem{}
Zel'dovich, Ya.B., Guseynov, O.H., 1966, ApJ 144, 840

\end{thebibliography}
\end{document}